\documentclass[aps,prl,reprint]{revtex4-1}
\usepackage{graphicx}  
\usepackage{epstopdf}
\usepackage{bm}        
\usepackage{amssymb} 
\usepackage{amsmath} 
\usepackage{stmaryrd}
\usepackage{color}
\usepackage{changes}

\begin{document}
\title{Depinning dynamics of crack fronts}
\author{Julien Chopin}
\email{julien.chopin@espci.fr}
\affiliation{Gulliver UMR 7083, CNRS - ESPCI ParisTech,
PSL  Research  University, Paris,  France}
\affiliation{Instititut Jean le Rond d'Alembert UMR 7190, Sorbonne Universit{\'e}s, CNRS - UPMC, Paris, France}
\affiliation{Instituto de F\'isica,  Universidade Federal da Bahia, Salvador-BA 40170-115, Brazil}
\author{Aditya Bhaskar}\author{Atharv Jog}
\author{Laurent Ponson}
\email{laurent.ponson@upmc.fr}
\affiliation{Instititut Jean le Rond d'Alembert UMR 7190, Sorbonne Universit{\'e}s, CNRS - UPMC, Paris, France}

\newcommand{\laurent}[1]{\textcolor{red}{#1}}
\newcommand{\julien}[1]{\textcolor{blue}{#1}
}\newcommand{\aditya}[1]{\textcolor{green}{#1}}
\newcommand{\vo}{v_\mathrm{0}}
\newcommand{\vm}{v_\mathrm{m}}
\newcommand{\vdep}{v_\mathrm{dep}}
\newcommand{\Gc}{G_\mathrm{c}}
\newcommand{\Gcd}{G_\mathrm{c1}}
\newcommand{\Gco}{G_\mathrm{c0}}
\newcommand{\Ep}{E_\mathrm{p}}
\newcommand{\Es}{E_\mathrm{s}}
\newcommand{\wo}{w_\mathrm{0}}
\newcommand{\fo}{\delta f_\mathrm{0}}
\newcommand{\dfs}{\delta f_\mathrm{s}}
\newcommand{\atan}{\mathrm{atan}}
\begin{abstract}
We investigate experimentally and theoretically the dynamics of a crack front during the micro-instabilities taking place in heterogeneous materials between two successive equilibrium positions. We focus specifically on the spatio-temporal evolution of the front, as it relaxes to a straight configuration, after depinning from a single obstacle of controlled strength and size. We show that this depinning dynamics is not controlled by inertia, but instead, by the rate dependency of the dissipative mechanisms taking place within the fracture process zone. This implies that the crack speed fluctuations around its average value $\vm$ can be predicted from an overdamped equation of motion $(v-\vm)/v_0 = (G-\Gc(\vm))/\Gc(\vm)$ involving the characteristic material speed $v_0 = \Gc(\vm)/\Gc'(\vm)$ that emerges from the variation of fracture energy with crack speed. Our findings pave the way to a quantitative description of the critical depinning dynamics of cracks in disordered solids and open up new perspectives for the prediction of the effective failure properties of heterogeneous materials.
\end{abstract}
\maketitle
Woods, nacre, bones or rationally designed artificial materials, are all heterogeneous solids, with mechanical properties far exceeding those of their constitutive components. Understanding the role of microscale heterogeneities on the macroscale fracture behavior of solids still remains a query. This becomes especially relevant now, as rapid developments in microfabrication techniques allow the tailoring of microstructures at ever smaller scales, yielding new types of composites, known as meta-materials, with unprecedented mechanical properties~\cite{Florijn,Leonard,blees,bertoldi2010negative,silverberg2014using,lechenault2014mechanical}. Recently, significant progresses were made for weakly heterogeneous brittle solids where models describing a crack front as a deformed interface pinned by tough obstacles have been successfully applied~\cite{Gao,Dalmas2,Patinet,Xia3,Vasoya2}. The homogenized fracture properties can be computed exactly within the so-called weak pinning limit~\cite{Roux4}, where the elastic energy release rate $G$ balances the fracture energy $G_\mathrm{c}$ at any time and any position along the front. This approach holds for weak variations of toughness along the propagation direction. The crack evolution is then smooth and can be properly approximated by a continuous succession of equilibrium front configurations~\cite{Hossain,Vasoya4}. This approach was successfully used to design weakly heterogeneous systems with improved and new macroscopic failure properties~\cite{Xia,Ghatak2,Kammer,Barthelat3}.

However, most natural and engineered materials have a microstructure composed of discontinuous heterogeneities which cannot be described within the weak pinning regime. The strong pinning regime that predominates for large toughness gradients challenges standard homogenization approaches. Crack propagation is not quasi-static but proceeds by intermittent and local micro-instabilities. Further, for a disordered distribution of obstacles, crack growth takes place close to the so-called depinning critical transition~\cite{Bonamy6,Alava2,Ponson19}, so that the crack front dynamics is dominated by avalanches spanning over a large range of length and time scales~\cite{Schmittbuhl4,Maloy2,Maloy3,Bonamy5,chopin2015morphology}. The precise understanding of the front evolution during these rapid events is a prerequisite to predict and further, control the fracture energy of heterogeneous solids. Beyond fracture, the behavior of driven disordered mechanical systems with long-range interactions is still an open question whose tremendous difficulty resides in the subtle interplay between fast, localized, depinning events and larger macroscopic avalanches forming a complex energetic landscape composed of many metastable states~\cite{Laurson2,Tallakstad,Ponson20}.
\begin{figure}
    \centering
    \includegraphics[width=8.5cm]{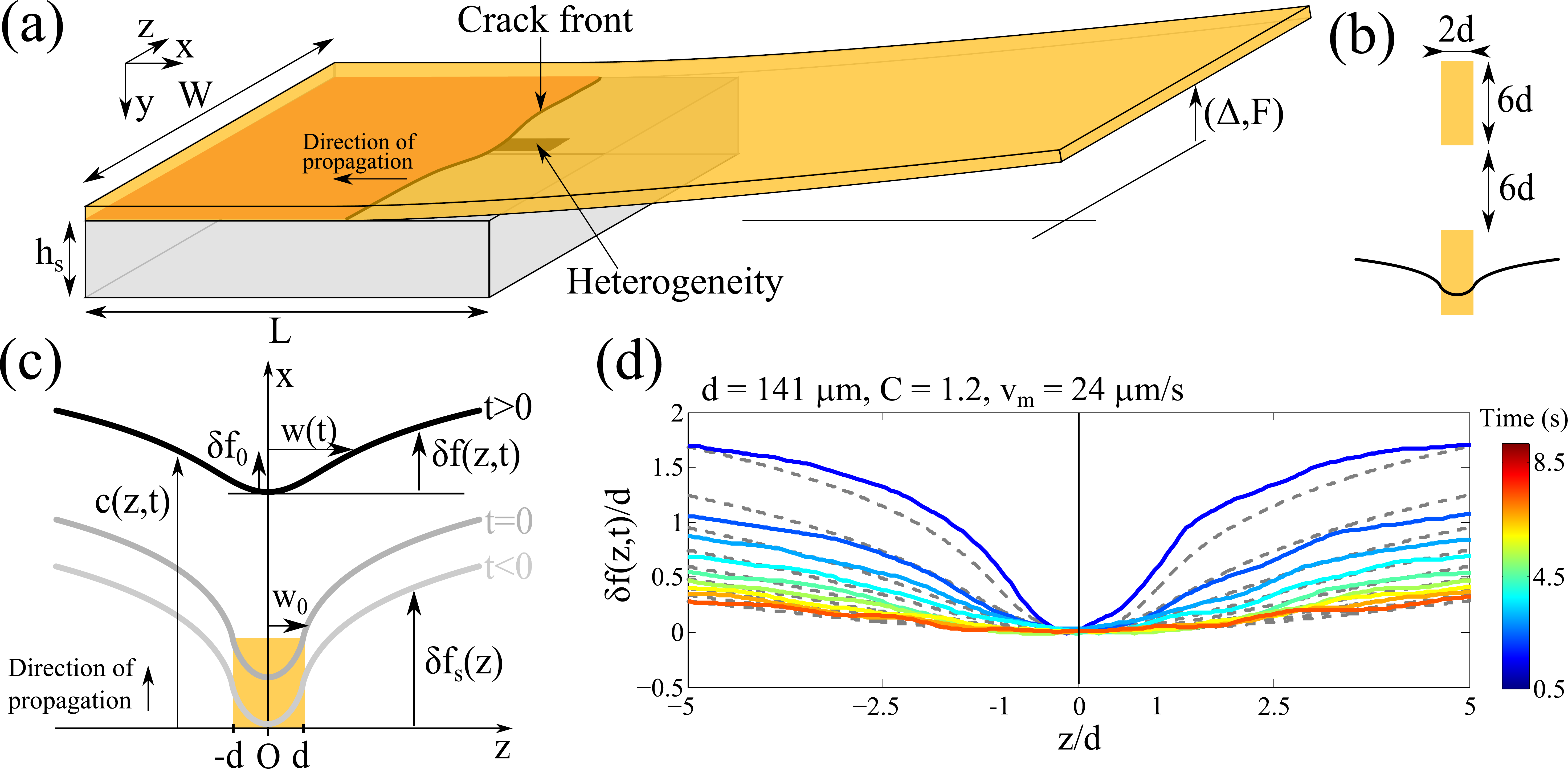}
    \caption{(a) Schematics of the experimental setup showing an interfacial crack front pinned by a heterogeneity. (b) Geometry of the rectangular obstacles of larger toughness. (c) Crack front positions in the stationary regime  $\dfs$ for ($t<0$) and during relaxation $c(z,t$ for $t>0$. $\delta f(z,t) = c(z,t)-c(0,t)$, $\delta f_0 = \delta f(d,0)$ and $w(t)$ are the crack front fluctuation, the characteristic height and half-width of the pinned region, respectively. 
    (d) Sequence of crack profiles after depinning from an obstacle 
    with theoretical prediction (dotted lines) with $\vo = 76\,\mu$m$\cdot$s$^{-1}$ as a unique fitting parameter
    }
    \label{Fig1}
\end{figure}

In this letter, we address experimentally and theoretically the basic problem of the interaction of a crack front with a tough obstacle in the strong pinning regime. In our experiment, a planar crack is driven at a constant speed over a tough region of finite length along the propagation direction, triggering a depinning instability between two well-defined metastable states. The size and strength of the obstacles are fully controlled and adjusted using our patterning technique. The sample allows in-situ visualization of the crack front dynamics which is resolved spatially and temporally. After normalizing all the length scales by the obstacle width $d$, we show that the relaxation dynamics follows a universal law which only depends on $v_0$, the crack speed at depinning for an obstacle of unit strength. $v_0$ is also found to vary linearly with the crack speed $v_\mathrm{m}$ imposed prior depinning by the loading rate. Next, we develop a theoretical model based on linear elastic fracture mechanics to quantitatively capture the observed behavior. Here, inertial effects can be neglected as the crack speed remains several orders of magnitude lower than the wave speed. Instead, we take into account the {\it rate dependency} of the fracture energy to quantitatively capture the effect of crack speed on the dissipative mechanisms taking place within the process zone. 
Thus, unlike perfectly brittle solids, crack may propagate at finite speed in dissipative materials as the elastic energy release rate may be constantly balanced by a rate dependent fracture energy.
Linearizing the equation of motion around $v_\mathrm{m}$, we obtain an analytical solution for the depinning of a crack from a single obstacle that is shown to capture quantitatively all our experimental observations. The implications of our results on the energy dissipated during fast fracture events and the fracture behavior of materials with randomly distributed obstacles are discussed in the final part of our paper.

We start by describing our experimental setup. A $5~\mathrm{mm}$ plate made of Polymethylmethacrylate (PMMA, Young modulus $\Ep = 1.8~\mathrm{GPa}$) with a heterogeneous coating is detached from a thick elastomer block using the beam cantilever geometry shown in Fig.~\ref{Fig1}(a). A vertical upward point like force is exerted at the extremity of the PMMA plate by means of a string connected to a mechanical testing machine allowing to impose the deflection speed. The elastomer is a crosslinked PolyDiMethylSiloxane (PDMS Sylgard184, Dow Corning) with a much lower Young modulus $\Es = 1.5~\mathrm{MPa}$ than PMMA and a Poisson's ratio $\nu_\mathrm{s} \simeq 0.5$. It is prepared by mixing an oligomer together with a silicon oil and degased for 2 hours under mild vacuum. It is then cured in an oven at 75$^{o}$C for at least 2 h. The resulting crosslinked PDMS block of size $W \times L = 50 \times 80~\mathrm{mm}^2$ with thickness $h_\mathrm{s} = 20~\mathrm{mm}$ is then demoulded. The crack is driven at an average speed $v_\mathrm{m}$ in the range $5 - 100~\mu\mathrm{m/s}$ that is set by the deflection rate imposed by the testing machine.

Taking inspiration from the experiments of Xia~{\it et al.}~\cite{Xia,Xia3}, we control the local fracture properties of the interface by printing obstacles on a commercial transparency, taking advantage of the high toughness $\Gcd$ of the printed regions on PDMS compared to the neat one noted $\Gco$. Unlike $\Gcd$ which does not show significant variations with the crack speed $v$, $\Gco$ is found to increase as $v^\gamma$ where $\gamma  = 0.37 \pm 0.05$~\footnote{See Supplemental Material for the toughness characterization and the detailed calculation of the depinning dynamics}. As a consequence, the contrast $C = (\Gcd - \Gco)/\Gco$ can be varied by exploring different crack speeds. As shown in Fig.~\ref{Fig1}(b), rectangles of width $2\, d$ and length $6\, d$ are aligned along the propagation direction where $d$ is varied between $0.1$ and $0.5~\mathrm{mm}$. A spacing of $6\, d$ between two successive obstacles is chosen to allow a complete relaxation of the front before it reaches the next obstacle. The transparency is then bonded onto the PMMA plate by means of a double-sided adhesive tape, the heterogeneous side faced up. Finally, a thin liquid film of PDMS is laid on the substrate before bringing the coated PMMA plate in contact allowing an intimate bonding between materials after curing at 40$^{o}$C for 48h. 

The transparency of the materials used in our setup is exploited to visualize the front geometry and its evolution as it interacts with the obstacle. Images of $3900 \times 2600$ pixels are taken normal to the mean fracture plane by a CCD camera through a semi-transparent mirror oriented at 45$^{o}$. An LED panel is placed horizontally beneath the sample to increase the contrast between the bonded and unbonded regions of the interface. A home-made algorithm extracts then the crack position $c(z,t)$ for each image taken at time $t$ where the depinning onset defines $t=0$ (see Fig.~\ref{Fig1}(c)). The front deformation is defined as $\delta f(z,t) = c(z,t) - c(0,t)$. An acquisition rate of $10~\mathrm{Hz}$ allows resolving in detail the front evolution during the depinning regime.

In a typical experiment, the front propagates initially in a homogeneous interface as a straight line. When crossing the obstacle, the profile gradually deforms until reaching a stationary shape composed of a pinned region of amplitude $\delta f_0(C,d)$ and logarithmic tails $\dfs(z)\, {\simeq}\, 2 \delta f_0(C,d) \log(|z|/d)$ for $|z| \gg d$. For weak obstacles, $\delta f_0(C,d)$ varies linearly with $C$ but non-linearities appear when $C$ is finite, yielding $\delta f_0(C,d) = d\,C/\pi(1 - C/2 + C^2/6)$~\cite{Vasoya2,chopin2011crack,Dalmas2,Dalmas2,Patinet,Vasoya}. When reaching the end of the obstacle, the crack front is suddenly out-of-equilibrium as the deformed profile is not stable in a homogeneous interface. We observe a fast motion of the pinned region and a slower motion of the remote part resulting in a relaxation towards a straight configuration. This behavior is reminiscent of avalanches which are sudden fracture events observed between metastable configurations in fully disordered materials driven close to the so-called depinning transition~\cite{Maloy2,Bonamy5,Laurson2}. We will see later that both phenomena are actually closely related.

\begin{figure}
    \centering
    \includegraphics[width=8.5cm]{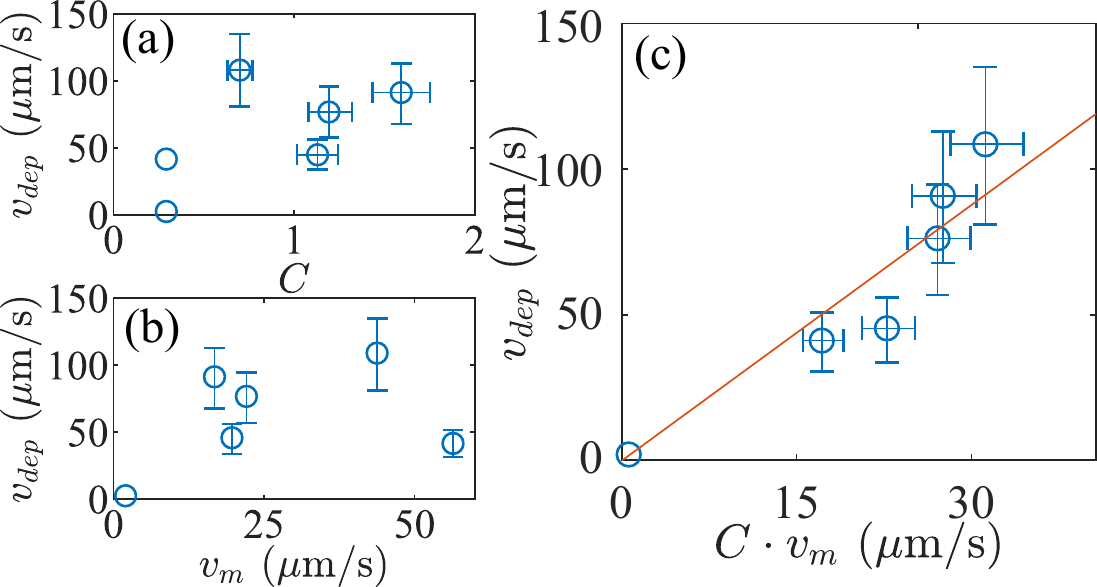}
    \caption{Depinning velocity $\vdep$ defined as the jump in crack speed as the front escapes from the obstacle varying (a) $C$ and (b) $v_m$. (c) $\vdep$ increases linearly with $C v_\mathrm{m}$.}
    \label{Fig2}
\end{figure}

We first focus on the initial dynamics of the instability measuring the depinning velocity $\vdep$ defined as $\vdep = dc_{|z|<d}/dt |_{t=0^+} - \vm$ where $c_{|z|<d}$ is the front position averaged over $|z|<d$. While most experimental and numerical studies only report averaged quantities such as avalanches duration and size, here we have access to the entire dynamics. We found that $\vdep$ is not uniquely determined by either $v_m$ or $C$ as indicated by the non-monotonous behaviors shown in Fig.~\ref{Fig2}(a) and (b). However, in Fig.~\ref{Fig2}(c), we show that $\vdep$ is linearly depending on $C \vm$ as revealed by the good collapse of the data onto a line of slope $\vdep/C \vm = \vo / \vm = 3.1$ where $\vo = \vdep/C$ is the depinning velocity for an obstacle of unit strength.

\begin{figure}
    \centering
    \includegraphics[width=5.5cm]{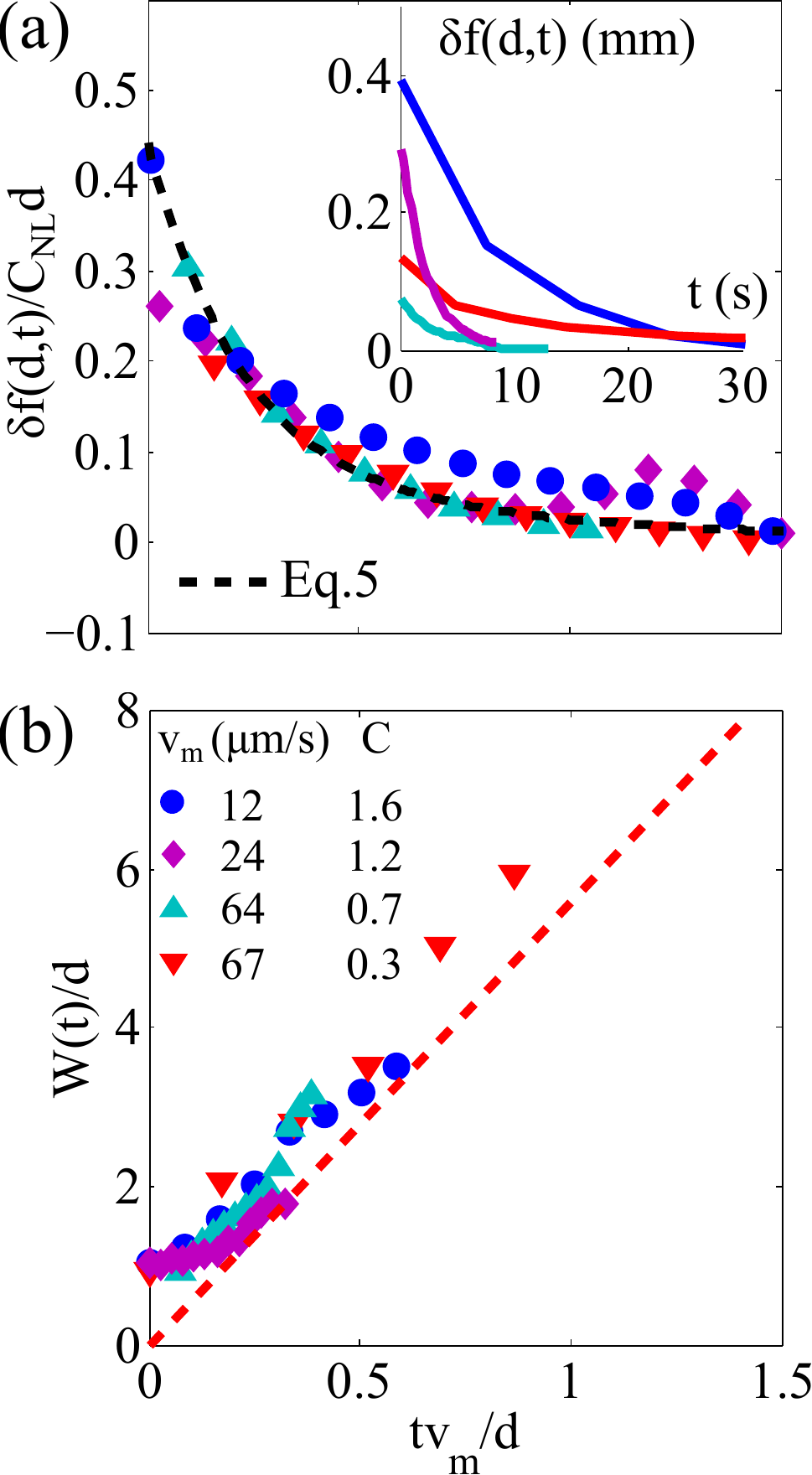}
    \caption{
    (a) Relaxation of the normalized deformation amplitude with theoretical prediction (dashed line). Inset : data before normalization. (b) Spreading of the deformation width $w(t)$ and theoretical prediction (dashed line).}
    \label{Fig3}
\end{figure}
 
Next, we address the relaxation dynamics at longer times beyond the onset of instability. We first measure the amplitude $\delta f(d,t)$ of the front deformation, and its evolution during depinning (see Fig.~\ref{Fig1}(c)). As shown in the inset of Fig.~\ref{Fig3}(a), we observe that $\delta f(d,t)$ relaxes towards zero at a rate strongly depending on $\vm$ and $C$. However, we found a good collapse of the relaxation profiles by normalizing $\delta f(d,t)$ and $t$ by $C\times d$ and $d/v_m$, respectively. These rescalings are found to be also relevant for the evolution of the half-width $w(t)$ of the perturbation, where $w(t)$ is defined from the relation $\delta f(w(t),t) = \dfs(d)$ (see Fig.~\ref{Fig1}(c)). Here, $w(t)$ quantifies the lateral spreading of the perturbation through time. As shown in Fig.~\ref{Fig3}(b), we also find a good collapse of the data normalizing $w(t)$ by $d$. Further, after a short transient, the width is found to grow linearly with time, following $w(t) = 5.7 v_m t$. \\ 

To explain quantitatively the observed dynamics, we develop a model within the framework of Linear Elastic Fracture Mechanics (LEFM) including a physically based dissipation mechanism to account for the viscoelastic dissipation in the process zone (PZ). Imposing that the energy release rate is balanced by the dissipated work within the PZ, the equation controlling crack evolution reads
\begin{equation}
    G[c(z,t)] = G_c[c(z,t),v(z,t)] \,.
    \label{Eq_Motion}
\end{equation}
Here, $G_c$ not only depends on the crack configuration $c(z,t)$ resulting from the interaction of the front with the obstacle but also on the local speed $v(z,t) = \partial{c}(z,t)/\partial{t}$ owing to the rate dependency of the dissipation. A first-order perturbation of Eq.~\eqref{Eq_Motion} around the mean front position $v_\mathrm{m} t$ yields  $\delta G[\delta c] = \frac{\partial\Gc (\vm)}{\partial v}\delta v$ where $\delta c(z,t) = c(z,t) - v_\mathrm{m} t$ and $\delta v(z,t) = v(z,t) - v_\mathrm{m}$. The l.h.s. term corresponds to a non-local elastic restoring force~\cite{Rice4} while the r.h.s term represents a local friction term increasing linearly with $v$. Terms such as $\frac{\partial \Gc}{\partial c}$ are not relevant since depinning occurs in a homogeneous region of the interface. The fracture toughness is taken in the form of $G_c = G_c^0 (v/v_c)^{\gamma}$, where $G_c^0$, $\gamma$ and $v_c$ are material parameters characterizing the dissipation mechanisms taking place in the process zone. Upon linearization of $G_\mathrm{c}$ around the macroscopic driving velocity $\vm$ in a slow propagation regime $\vm \ll v_c$, we obtain
\begin{equation}
   \frac{1}{G_c^0} \frac{\delta G_c}{\delta v} \equiv \frac{1}{\vo} = \frac{\gamma}{\vm} .
    \label{Gc_lin}
\end{equation}
Here we take $\gamma = 1/3$ in agreement with the value measured for the neat regions of the interface between the PDMS substrate and the cantilever~\cite{Note1}. 
Using the expression of $\delta G$ derived for an interfacial crack between an incompressible substrate and a much stiffer material~\cite{Pindra}, we obtain the equation of motion
\begin{equation}
    \frac{1}{\vo} \frac{\partial \delta c(z,t)}{\partial t}= \frac{1}{\pi} PV \int_{-\infty}^{+\infty} dz' \frac{\delta c (z',t) - \delta c (z ,t)}{(z'-z)^2}.
    \label{eq:motion}
\end{equation}
It is noteworthy that Eq.~\ref{eq:motion} is formally equivalent to the linear order to the equation of motion of a contact line of Newtonian fluids partially wetting a solid surface~\cite{deGennesWettingRMP,katzav2007roughness}. The steady-state pinned profile is taken as initial condition. The equation of propagation can be solved exactly, yielding
\begin{equation}
    \frac{\pi}{C}\frac{\delta v}{\vo}= \arctan \left( \frac{z+d}{\vo t} \right) -\arctan \left(\frac{z-d}{\vo t} \right).
    \label{Eq_v}
\end{equation}
From Eq.~\eqref{Eq_v}, we obtain the depinning velocity $v_\mathrm{dep} = C v_0 \approx 3 C v_\mathrm{m}$ which is in very good agreement with the experimental data of Fig.~\ref{Fig2}(c) (solid line). Note that the existence of a characteristic depinning speed $v_0$ emerging from the kinetic law $G_\mathrm{c}(\vm)$ had already been noticed by Kolvin~{\it et al.}~\cite{kolvin2017nonlinear} during the formation and death of microbranch that effectively acts as localized pinning point for the crack front.

\begin{figure}
    \centering
    \includegraphics[width=5.5cm]{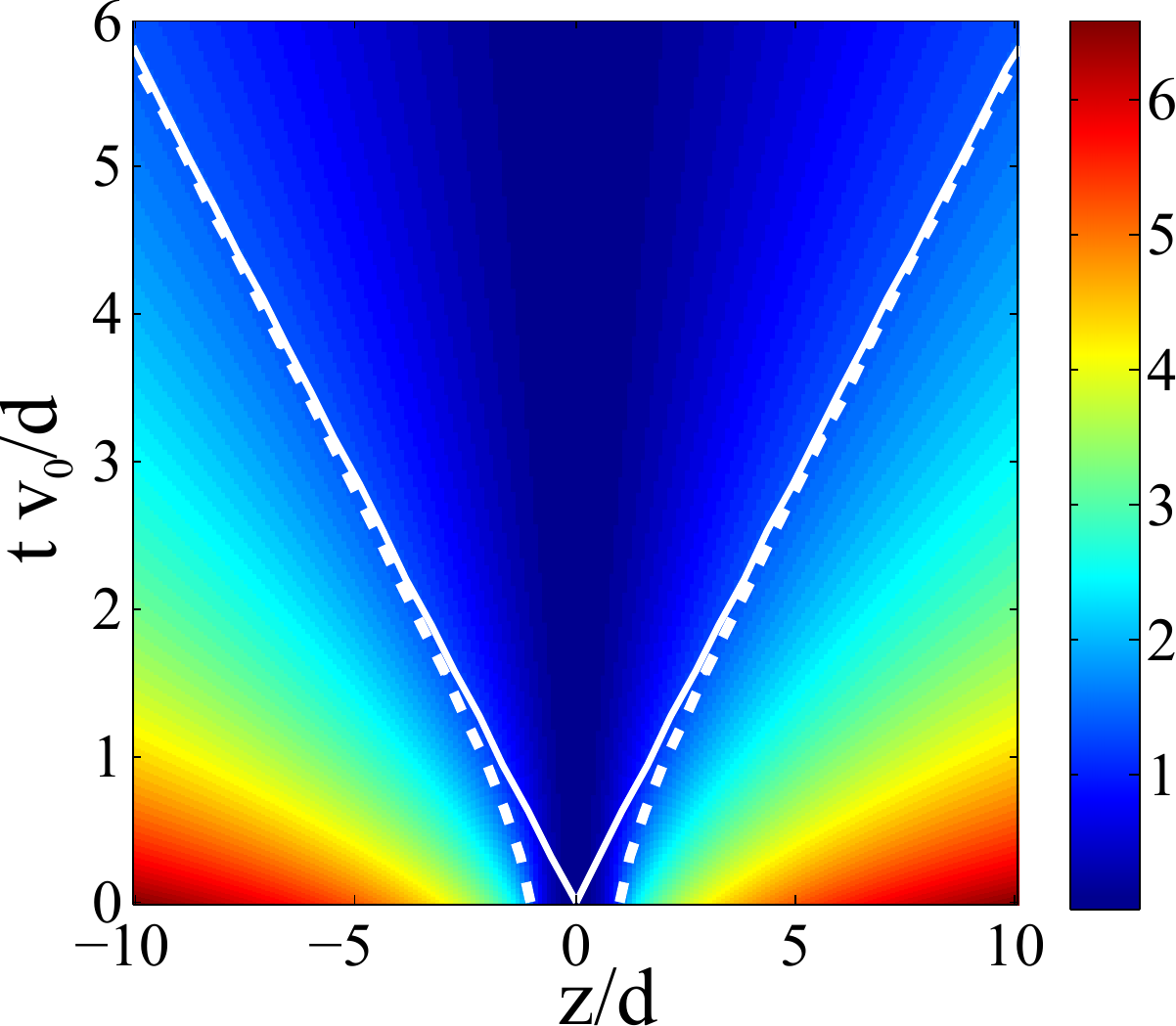}
    \caption{Analytical spatio-temporal map of the front deformation $\delta f$ during depinning. The contour line for $\delta f=\delta f_0$ obtained using the exact solution (dashed line) and the asymptotics solution for $|z| \gg d$ (Eq.~\ref{Eq_w}, solid line) are shown.}
    \label{Fig4}
\end{figure}

The speed profile can then be readily integrated to provide the general form of the front profile. Fig.~\ref{Fig4} shows a spatio-temporal map of $\delta f(z,t)$ where the entire relaxation to a straight configuration can be observed. To avoid cumbersome equations, we will just give analytical expressions of $\delta f(z,t)$ in some limits which are useful to interpret our experimental data~\cite{Note1}. For $z=d$, we obtain
\begin{equation}
\begin{array}{ll}
\displaystyle    \frac{\pi \delta f(d,t)}{Cd} &=  \frac{t\,v_0}{d} \left[ \arctan \left ( \frac{2d}{t\,v_0} \right ) -2 \arctan \left ( \frac{2d}{t\,v_0} \right ) \right ] \\
 &\displaystyle +\ln \left [ \frac{ 4+\left(\frac{t\,v_0}{d}\right)^2}{1+\left(\frac{t\,v_0}{d}\right)^2} \right ]
\end{array}
    \label{Eq_h}
\end{equation}
Eq.~\ref{Eq_h} is in good agreement with the experimental data of Fig.~\ref{Fig3}(b) (dashed line) provided that the amplitude of the perturbation is normalized by $C_{NL} = C(1-C/2+C/6)$ which is justified owing to the rather large contrasts values explored in our experiments. In the limit $|z| \gg d$, we obtain $\delta f(z,t) = \displaystyle \ln \left(1 + \left(z /(t v_0)\right)^2 \right)$ which generalizes the model first obtained by Marsh \& Cazabat~\cite{marsh1993dynamics} for the depinning of a contact line~\cite{Note1}. Thus, we obtain
\begin{equation}
    w(t) \simeq \sqrt{3}v_0 t = (\sqrt{3}/\gamma) v_m t .
    \label{Eq_w}
\end{equation}
 As shown in Fig.~\ref{Fig4}, the linear spreading of the perturbation provides a good approximation even at relatively short time. Eq.~\eqref{Eq_w} captures well the experimental observations of Fig.~\ref{Fig3}(b).

To summarize, our study of the depinning of a brittle crack from a single obstacle reveals a characteristic velocity $v_0$ that sets the relaxation time $\lambda/v_0$ of the front perturbations of wavelength $\lambda$. This characteristic speed that emerges from the crack growth law $G_\mathrm{c}(v_\mathrm{m})$ allows us to derive an overdamped equation of motion $(v-\vm)/\vo = (G - \Gc(\vm))/\Gc(\vm)$ that was shown to capture quantitatively the crack front evolution during depinning as observed in our experiments.

Following are the implications of these findings. First, it sheds light on the nature of the dissipation accompanying avalanches in failure of heterogeneous solids. During an avalanche, the depinning region of the front reaches the speed $v_0$ that may be much larger than the average crack speed  $v_\mathrm{m}$. Owing to the increase of the fracture energy with crack speed and the continuity of the elastic energy at the onset of depinning, the dissipation rate during an avalanche is close to the toughness of the impurities, leading to an additional dissipation that reduces to $\simeq C^2 \, G_\mathrm{c0} d^2$ per heterogeneity for the case of a periodic array of obstacles. For disordered distributions, in the strong pinning regime where the front motion consists of a succession of avalanches, we then expect the energy dissipated by unit fractured surface to be significantly larger than the matrix toughness, and closer to the obstacle fracture energy, even for relatively low obstacle density. The proposed crack evolution equation that is amenable to the exploration of more complex toughness landscape embedding multiple obstacles predict the total energy dissipated, including the contribution due to depinning instabilities, and so can serve as a tool for the design of patterned interfaces with improved mechanical performance.

Secondly, our findings allow to address a long standing question about the failure of disordered solids and its relationship with critical phenomena. For randomly distributed obstacles, cracks exhibit a jerky dynamics characterized by universal scaling laws that were shown to be reminiscent of the so-called depinning transition of an elastic interface driven in a random medium~\cite{Schmittbuhl4,Bonamy5,Ponson14}. However, the control parameter that sets the distance of the system to the critical point was not identified yet, in particular under displacement controlled conditions where the front velocity $v_\mathrm{m}$ is imposed. From the description of the crack dynamics during unstable events brought by this study, this can now be achieved through the comparison of the driving velocity $v_\mathrm{m}$ with the characteristic speed $v_0$ of the avalanches, leading to the control parameter $\delta = v_\mathrm{m}/v_0$. As expected for dynamical phase transition, this parameter controls the crack front behavior, like the correlation time of the speed fluctuations that was recently shown to diverge as $1/\delta$ ~\cite{Ponson20,Tallakstad}. Interestingly, $v_0$ may not be independent of $v_\mathrm{m}$. For many material systems like the one considered in this study, $G_\mathrm{c}$ increases as a power law of $v_\mathrm{m}$ so that $v_0 = v_\mathrm{m}/\gamma$ (see Eq.~\eqref{Gc_lin}). As a result, the control parameter may often take a fixed value $\delta = \gamma$, explaining why the crack response is independent of its average speed over several orders of magnitude~\cite{Maloy3,Plouraboue2}. Overall, the introduction of the parameter $\delta = v_\mathrm{m}/v_0$ that controls the critical behavior of fracturing material opens new perspective for the quantitative description of fracture in terms of depinning transition.

\end{document}